\documentclass[aps,twocolumn]{revtex4}%
\usepackage{amsfonts}
\usepackage{amsmath}
\usepackage{amssymb}
\usepackage{graphicx}
\begin{document}

\title{Massless QED$_{3}$ with explicit fermions}
\author{Dean Lee}
\email[]{dean_lee@ncsu.edu}
%\thanks{}

\author{Pieter Maris}
\email[]{pmaris@unity.ncsu.edu}
%\thanks{}
\affiliation{Department of Physics,
North Carolina State University, 
Raleigh,  NC 27695-8202}

\date{\today}

\begin{abstract}
We study dynamical mass generation in QED in $(2+1)$ dimensions using
Hamiltonian lattice methods.  We use staggered fermions, and perform
simulations with explicit dynamical fermions in the chiral limit.  We
demonstrate that a recently developed method to reduce the fermion
sign problem can successfully be applied to this problem.  Our results
are in agreement with both the strong coupling expansion and with
Euclidean lattice simulations.
\end{abstract}

%\pacs{11.10.St, 11.30.Rd, 12.38.Lg, 14.40.Aq}

%\keywords{}

\maketitle

%%%%%%%%%%%%%%%%%%%%%%%%%%%%%%%%%%%%%%%%%%%%%%%%%%%%%%%%%%%%%%%%%%%%%%%%%%%%%
\section{Introduction}

Quantum electrodynamics in $(2+1)$ dimensions (QED$_{3}$) is a theory
which shares a number of important features with quantum
chromodynamics in $3+1$ dimensions (QCD) such as dynamical mass
generation~\cite{pisa,Burkitt:1987nx,Hands:1989mv,Dagotto,dcsb,pennington,Luo:kk,Azcoiti,Hamer:1997bf,Hands:2002dv}
and confinement~\cite{conf}.  Since QED$_{3}$ is super-renormalizable
and has fewer degrees of freedom than QCD, it serves as a valuable
laboratory in which to test new methods and ideas related to these
phenomena.  Aside from its role as a testing ground for QCD, however,
QED$_{3}$ in itself plays an important role in solid state physics and
in particular high-$T_{c}$
superconductivity~\cite{Dorey:1991kp,Farakos:1997qi}.  Recently
several studies have pursued a new theoretical approach to cuprate
superconductors~\cite{balents,franz,superTc} in which one describes
the phase transition in the reverse direction, starting from the
superconducting state.  In this picture the antiferromagnetic phase,
for example, corresponds to spontaneous chiral symmetry breaking of
massless two-flavor QED$_{3}$.  But there are also several other
phases, and the large chiral manifold of degenerate states explains
the complexity of the phase diagram.

There are extensive studies using the Dyson--Schwinger
equations~\cite{dcsb} suggesting that chiral symmetry in QED$_{3}$ is
dynamically broken if the number of fermion flavors is smaller than
some critical number $N_{c}\sim3.3$.  However, the scale of this
symmetry breaking (i.e. the magnitude of the chiral condensate) is
extremely small, and there are also studies~\cite{pennington}
suggesting that chiral symmetry is broken for all number of fermion
flavors.  Quenched lattice simulations have shown clear signs of
chiral symmetry breaking, with a condensate
$\langle\bar{\psi}\psi\rangle \sim 5 \times 10^{-3}$ in units of
$e^4$, the dimensionful coupling constant~\cite{Hands:1989mv}.  The
situation for dynamical fermions however is not so clear, especially
for an odd number of flavors.  There have been Euclidean lattice
studies in both compact~\cite{Burkitt:1987nx} and
non-compact~\cite{Dagotto} formalisms with different numbers of
flavors, all suggesting a very small condensate.  The most recent
Euclidean lattice study of two-flavor non-compact QED$_3$ suggests an
upper bound for the condensate of $\sim 5 \times
10^{-5}$~\cite{Hands:2002dv}, using large lattices.

On the other hand, Hamiltonian lattice studies of QED$_3$ with one
fermion flavor have suggested a rather large value for the chiral
condensate.  These studies were based on the strong coupling
expansion~\cite{Hamer:1997bf} and variational coupled cluster
expansion~\cite{Luo:kk}.  The obtained condensate was $\sim
0.28$~\cite{Hamer:1997bf}, significantly larger than both quenched and
two-flavor Euclidean lattice results and about two orders of magnitude
larger than the Dyson--Schwinger results for one flavor QED$_3$.

In this paper we study chiral symmetry breaking in one-flavor massless
QED$_{3}$.  To our knowledge our analysis~\cite{Maris:2002sc}
represents the first non-perturbative simulation of lattice gauge
theory in more than one spatial dimension with explicit fermions.  By
explicit fermions, we mean that fermions are not integrated out to
yield determinants of the Dirac operator.  In the simulation presented
here, fermion dynamics are sampled explicitly using fermion worldlines
in a gauge-field dependent Hamiltonian.  From a theoretical point of
view, this is an ideal framework in which to address the fermion
structure of the ground state wavefunction.  From a computational
point of view, however, the approach presents profound difficulties
such as the fermion sign problem and complex phase fluctuations due to
the gauge field, both of which scale exponentially with the volume of
the system.  Therefore it is not likely that this approach would be
possible for QCD in the near future.  However, we do find that by
employing the recently developed \textit{zone method}~\cite{dean02},
we can control sign and phase problems sufficiently to study chiral
symmetry breaking in massless QED$_{3}$ on relatively small spatial
lattices.

In our study, we find that in the strong coupling region, $y<1$, our
results agree very well with the strong coupling
expansion~\cite{Hamer:1997bf}.  However, the agreement between the
strong coupling expansion and our simulations breaks down around $y
\sim 1$, and for $y>1$ we see a dramatic decrease in the size of the
condensate.  These results are in agreement with Euclidean lattice
simulations using staggered fermions~\cite{Burkitt:1987nx}, suggesting
a very small condensate in the continuum limit, $y\to\infty$.

\section{QED in $(2+1)$ dimensions}

QED$_{3}$ is a super-renormalizable theory, with a dimensionful
coupling: $e^{2}$ has dimensions of mass.  This dimensionful parameter
plays a role similar to $\Lambda_{\mathrm{QCD}}$ in QCD.  In the chiral
limit, it also sets the energy scale.  We use 4-component spinors,
such that the fermion mass term is even under parity.  With one
massless fermion flavor, the Hamiltonian exhibits a global $U(2)$
\textquotedblleft chiral\textquotedblright\ symmetry.  A fermion mass
term breaks this symmetry to a $U(1)\times U(1)$ symmetry.  The
question is: is this chiral symmetry broken dynamically? The order
parameter for this symmetry breaking is the chiral condensate.

\subsection{Lattice Hamiltonian}

We start with the staggered fermion lattice Hamiltonian on an
$L_{1}\times L_{2}$ spatial lattice~\cite{Hamer:1997bf},
\begin{eqnarray}
  H_{\text{physical}} &=& \frac{g^{2}}{2a}\big(W_{E}+W_{B}+W_{F}\big) \,,
\label{eq:physH}
\end{eqnarray}
with
\begin{eqnarray}
W_{E} &=& \sum_{\vec{r},j}\big(E_{j}(\vec{r})\big)^{2} \,,\\
W_{B} &=& -y^{2}\sum_{\vec{r}}\big(  
	U_{p}(\vec{r})+U_{p}^{\dagger}(\vec{r})\big)  \,,\\
W_{F} &=&-\mu\sum_{\vec{r}}(-1)^{r_{1}+r_{2}}
	\chi^{\dagger}(\vec{r})\chi(\vec{r})
\nonumber\\ &&{}
  	+ y\sum_{\vec{r},j}\eta_{j}(\vec{r})\chi^{\dagger}(\vec{r})
	U_{j}(\vec{r})\chi(\vec{r} + \hat{\jmath}) + \text{h.c.} \,,
\end{eqnarray}
where $\eta_{1}(\vec{r})=(-1)^{r_{2}+1}$, $\eta_{2}(\vec{r})=1$,
$y=1/g^{2}$, and $U_{p}$ is the plaquette operator given by the
product of $U_{j}(\vec{r})$'s circuiting the spatial plaquette
anchored at $\vec{r}$,
\begin{eqnarray}
U_{p}(\vec{r})=U_{1}(\vec{r})U_{2}(\vec{r}+\hat{1})U_{1}^{\dagger}(\vec
{r}+\hat{2})U_{2}^{\dagger}(\vec{r}).
\end{eqnarray}
We use a dimensionless mass parameter $\mu=2m/e^{2}$ and a
dimensionless coupling constant $g^{2}=e^{2}a,$ where $a$ is the
lattice spacing.  We also make the Hamiltonian and time dimensionless:
the actual simulations are performed with the {\em dimensionless}
Hamiltonian
\begin{eqnarray}
  H &:=& \frac{2a}{g^2} H_{\text{physical}} = W_{E}+W_{B}+W_{F} \,,
\end{eqnarray}
in combination with the dimensionless time variable
\begin{eqnarray}
  t &:=& \frac{g^2}{2a} t_{\text{physical}} \,,
\end{eqnarray}
instead of the physical Hamiltonian, Eq.~(\ref{eq:physH}).  

We use the Dirac matrix representation
\[
\gamma_{0}=\left[
\begin{array}
[c]{cc}%
\sigma_{3} & 0\\
0 & -\sigma_{3}%
\end{array}
\right] \;,\;\;\gamma_{1,2}=\left[
\begin{array}
[c]{cc}%
i\sigma_{1,2} & 0\\
0 & -i\sigma_{1,2}%
\end{array}
\right] \; .
\]
Assuming that $L_{1}$ and $L_{2}$ are even, we stagger the fermion
components at the four sites of a $2\times2$ unit cell
\begin{eqnarray}
\chi_{1}(\vec{r}) &\sim&(2r_1,2r_2)\\
\chi_{2}(\vec{r}) &\sim&(2r_1+1,2r_2)\\
\chi_{3}(\vec{r}) &\sim&(2r_1+1,2r_2+1)\\
\chi_{4}(\vec{r}) &\sim&(2r_1,2r_2+1)\,.
\end{eqnarray}
In the continuum limit the staggered fermions correspond to one
flavor of a 4-component fermion~\cite{Hamer:1997bf,Burden:qb},
\begin{eqnarray}
\frac{i^{r_{1}+r_{2}}}{2\sqrt{2}a}\left[
\begin{array}
[c]{cccc}%
0 & -i & 0 & 1\\
1 & 0 & -i & 0\\
-i & 0 & 1 & 0\\
0 & 1 & 0 & -i
\end{array}
\right]  \left[
\begin{array}
[c]{c}%
\chi_{1}(\vec{r})\\
\chi_{2}(\vec{r})\\
\chi_{3}(\vec{r})\\
\chi_{4}(\vec{r})
\end{array}
\right]  \rightarrow\left[
\begin{array}
[c]{c}%
\psi_{1}(\vec{r})\\
\psi_{2}(\vec{r})\\
\psi_{3}(\vec{r})\\
\psi_{4}(\vec{r})
\end{array}
\right]  \,.
\end{eqnarray}

For the states in our physical Hilbert space we choose a basis that is
a tensor product of the gauge field and the fermion field degrees of
freedom.  For each gauge link field let us define the gauge field
basis,
\begin{eqnarray}
U_{j}(\vec{r}) |X_{j}(\vec{r})\rangle &=&
	e^{iX_{j}(\vec{r})} |X_{j}(\vec{r})\rangle \,,
\end{eqnarray}
where each $X_{j}$ is a real number in the interval $[0,2\pi)$.  We
let $|X\rangle $ be the tensor product of states
$|X_{j}(\vec{r})\rangle $ at each link,
\begin{eqnarray}
 | X \rangle &=& \bigotimes_{\vec{r},j}
	|X_{j}(\vec{r})\rangle \,.
\end{eqnarray}
Consider the Green's function
\begin{eqnarray}
G_{X',\alpha';X,\alpha}(\Delta t) &=&
	\Big(\langle \alpha'|\otimes\langle X'|\Big)
	e^{-H \Delta t}
	\Big(|X\rangle\otimes |\alpha\rangle \Big) \,,
\nonumber \\ {}
\label{eq:greens}
\end{eqnarray}
where $|X\rangle \otimes |\alpha\rangle$ and
$|X'\rangle\otimes|\alpha'\rangle $ are two states in our physical
Hilbert space, with $|\alpha\rangle $ and $|\alpha^{\prime}\rangle$
general fermion states.  If $\Delta t$ is small and
$X_{j}(\vec{r})\approx X'_{j}(\vec{r})$ for all $\vec{r}$ and $j$,
then
\begin{eqnarray}
\lefteqn{ G_{X',\alpha';X,\alpha}(\Delta t) }
\nonumber\\
&  \propto & \exp\left[  -\tfrac{1}{4\Delta t} \sum_{\vec{r},j} 
	\big(X_{j}(\vec{r})-X'_{j}(\vec{r})\big)^2
			-\Delta t \, W_{B}^{X}\right]  
\nonumber\\
&& {}\times \langle \alpha^{\prime}| 
	\exp\left[-\Delta t \,W_{F}^{X}\right]  
	|\alpha\rangle \,,
\label{eq:greenexp}
\end{eqnarray}
where
\begin{eqnarray}
W_{B}^{X} &=&  W_{B}\Big|_{U_{j}(\vec{r})=e^{iX_{j}(\vec{r})}} \,,\\
W_{F}^{X} &=&  W_{F}\Big|_{U_{j}(\vec{r})=e^{iX_{j}(\vec{r})}} \,.
\end{eqnarray}
We can evaluate $G_{X_f,\alpha_f;X_i,\alpha_i}(t)$ for general initial
and final states and arbitrary $t$ by breaking the exponential in
Eq.~(\ref{eq:greens}) into $N$ equal time steps and inserting a
complete set of states at each time step $\Delta t = t/N$.  If $\Delta
t$ is small, the sum over intermediate states is dominated by
consecutive states that are similar, i.e., the $n^{\text{th}}$ time
step is dominated by states which satisfy $X^{(n+1)}_{j}\approx
X^{(n)}_{j}$ and thus we can use Eq.~(\ref{eq:greenexp}) repeatedly.
If we let $X_{i}=X^{(0)}$, $X_{f}=X^{(N)}$, $\alpha_{i}=\alpha
^{(0)}$, and $\alpha_{f}=\alpha^{(N)}$, then
\begin{eqnarray}
G_{X_{f},\alpha_{f};X_{i},\alpha_{i}}(t) &=&
{\textstyle\sum\limits_{\substack{X^{(1)},...,X^{(N-1)}\\
			\alpha^{(1)},...,\alpha^{(N-1)}}}}
A^{X,\alpha} \; B^{X}
\end{eqnarray}
where
\begin{eqnarray}
A^{X,\alpha} &=&  \prod_{n=0}^{N-1}\big\langle \alpha^{(n+1)}\big| 
   e^{-\tfrac{t}{N}W_{F}^{X^{(n)}}}\big|\alpha^{(n)}\big\rangle \,,
\\
B^{X} &=& \prod_{n=0}^{N-1} \exp{\left[-\tfrac{N}{4t} \sum_{\vec{r},j}
	\big(X_{j}^{(n+1)}(\vec{r})-X_{j}^{(n)}(\vec{r})\big)^2\right]}
\nonumber \\ && \hphantom{\prod_{n=0}^{N-1} }
	\times \exp{\left[-\tfrac{t}{N}W_{B}^{X^{(n)}}\right]} \,.
\end{eqnarray}

In our simulation the gauge field configurations are updated using the
Metropolis algorithm.  For each new gauge configuration we compute
the evolution of the corresponding time-dependent Hamiltonian in the
space of fermionic states.  The sampling over fermion states is
performed using the worldline formalism~\cite{worldline}, which we now
briefly discuss.

\subsection{Worldlines}

At the $n^{\text{th}}$ time step we have an exponential operator of
the form
\begin{eqnarray}
S^{(n)} &=& e^{-\sum_{\vec{r}}\big(  H_{0}(\vec{r}) 
	+ H_{1}^{X^{(n)}}(\vec{r})
	+ H_{2}^{X^{(n)}}(\vec{r})\big)  } \, ,
\end{eqnarray}
where
\begin{eqnarray}
H_{0}(\vec{r}) &=& -\tfrac{N\mu(-1)^{r_{1}+r_{2}}}{t}
	\chi^{\dagger}(\vec{r})\chi(\vec{r})\\
H_{1}^{X^{(n)}}(\vec{r}) &=& \tfrac{y N \eta_{1}(\vec{r})
	e^{iX_{1}^{(n)}(\vec{r})}}{t}\chi^{\dagger}(\vec{r})
	\chi(\vec{r}+\hat{1}) + \text{h.c.}\\
H_{2}^{X^{(n)}}(\vec{r}) &=& \tfrac{y N \eta_{2}(\vec{r})
	e^{iX_{2}^{(n)}(\vec{r})}}{t}\chi^{\dagger}(\vec{r})
	\chi(\vec{r}+\hat{2}) + \text{h.c.}
\end{eqnarray}
In the following we use the shorthand \textquotedblleft
e\textquotedblright\ for even and \textquotedblleft
o\textquotedblright\ for odd values of $r_j$, $j=1,2$.  Let us break
up $S^{(n)}$ into a product of four terms,
\begin{eqnarray}
S^{(n)} &\approx& S_{2;\text{o}}^{(n)}S_{2;\text{e}}^{(n)}
	S_{1;\text{o}}^{(n)}S_{1;\text{e}}^{(n)} \;,
\end{eqnarray}
where
\begin{eqnarray}
	S_{j;\text{e/o}}^{(n)} &=& 
	\exp\left[-\sum_{r_{j;\text{e/o}}}
	\Big(  \frac{1}{4}H_{0}(\vec{r})
	+ H_{j}^{X^{(n)}}(\vec{r})\Big)\right] \,.
\end{eqnarray}
Each $S_{j;\text{e/o}}^{(n)}$ is the product of mutually commuting
operators which contain the interactions for an adjacent two-site
system.  With this decomposition of the time evolution operator, one
can trace out the worldline of any individual fermion.  
\begin{figure}[bht]
{\begin{center}
\includegraphics[width=5.5cm]{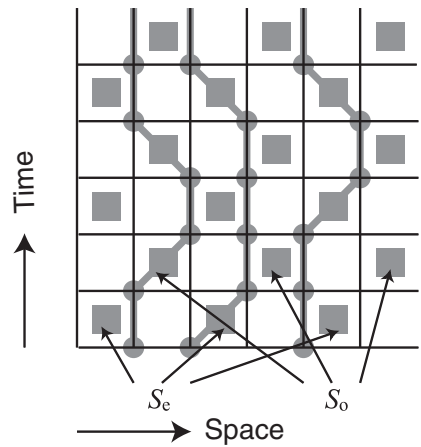}
\end{center}}
\caption{Sample worldline configuration of a system with one spatial
dimension. \label{fig:one}}
\end{figure}
In Fig.~\ref{fig:one} we have drawn the worldlines for a sample
worldline configuration.  For visual clarity the example we have drawn
is a simpler system with only one spatial dimension.  We have placed
shaded squares where the interactions $S_{j;\text{e/o}}^{(n)}$ occur.
In the case when two identical fermions enter the same shaded square
we use the convention that the worldlines run parallel and do not
cross.

\begin{figure}[tbh]
{\begin{center}
\includegraphics[width=5.5cm]{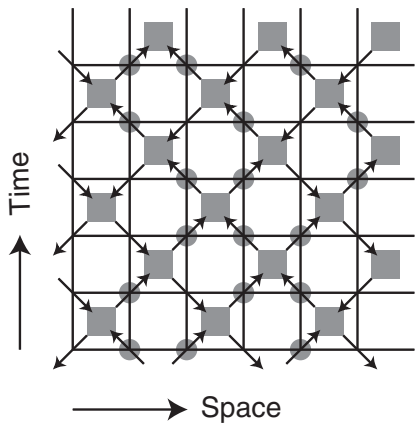}
\end{center}}
\caption{Upward/downward arrows are drawn at each occupied/unoccupied
site. \label{fig:two}}
\end{figure}

The sum over all worldline configurations is calculated with the help
of the loop algorithm~\cite{loop}.  At each occupied/unoccupied site,
we place an upward/downward pointing arrow as shown in
Fig.~\ref{fig:two}.  Due to fermion number conservation, the number of
arrows pointing into a shaded square equals the number of arrows
pointing out of the square.  As a consequence of this conservation
law, any valid worldline configuration can be generated from any other
worldline configuration by flipping arrows that form closed loops.
New Monte Carlo updates are therefore implemented by picking a random
closed loop and using the Metropolis condition to determine whether or
not to flip the loop.

\subsection{Measuring the chiral condensate}

For $\mu=0$, the staggered lattice formulation reduces the chiral
$U(2)$ symmetry to a discrete symmetry generated by a shift of one
lattice spacing.  To study chiral symmetry breaking on the lattice, we
calculate the lattice condensate $\langle\bar{\psi}\psi\rangle$ in the
chiral limit as a function of the lattice coupling $y$.  The lattice
condensate is related to the continuum condensate by the relation
\begin{eqnarray}
 y^{2}\langle\bar{\psi}\psi\rangle^{\text{lattice}}
	&=&\frac{1}{e^{4}}
	\langle\bar{\psi}\psi\rangle^{\text{continuum}}\;,
\end{eqnarray}
in the limit $y\rightarrow\infty$.  From here on
$\langle\bar{\psi}\psi\rangle$ will denote
$\langle\bar{\psi}\psi\rangle^{\text{lattice}}$.  We determine the
lattice condensate by computing the limit
\begin{eqnarray}
 \lim_{t\rightarrow\infty}
	\frac{\big(\langle\alpha|\otimes\langle F|\big)
	e^{-\frac{H t}{2}} {\cal O} e^{-\frac{H t}{2}}
	\big(|F\rangle \otimes|\alpha\rangle\big)}
	{\big(\langle\alpha|\otimes\langle F|\big)
	e^{-H t} \big(|F\rangle\otimes|\alpha\rangle\big)}\,,
\label{ratio}
\end{eqnarray}
where
\begin{eqnarray}
\cal{O} &=& -\tfrac{1}{L_{1}L_{2}} \sum_{\vec{r}}(-1)^{r_{1}+r_{2}}
	\chi^{\dagger}(\vec{r})\chi(\vec{r})\,.
\end{eqnarray}
The state $|F\rangle $ is a variational approximation to the gauge
field ground state,
\begin{eqnarray}
 |F\rangle &=& \int dX \; F(X) |X\rangle  \, ,
\end{eqnarray}
where
\begin{eqnarray}
 F(X)= e^{c\sum_{\vec{r}}\cos\left(
	X_{1}(\vec{r})+X_{2}(\vec{r}+\hat{1})
	-X_{1}(\vec{r}+\hat{2})-X_{2}(\vec{r})\right) }\,,
\end{eqnarray}
and $c$ is a real parameter we choose to optimize overlap with the
true gauge field ground state~\cite{green's}.  In our simulations we
have used $c=\frac{y}{4}$, which appears to work well for both small
and large $y$.  The state $|\alpha\rangle$ is the $y=0$ fermion ground
state for $\mu>0$, a configuration where even sites are occupied and
odd sites are unoccupied.  The essential characteristic of the trial
state $|F\rangle \otimes |\alpha\rangle$ is that it has non-zero
overlap with the physical vacuum.

In Eq.~(\ref{ratio}), for both numerator and denominator, the initial
quantum state is the same as the final state.  Any configuration of
fermion worldlines can therefore be regarded as a permutation of the
initial fermions.  Even permutations give a positive contribution
while odd permutations come with a minus sign.  Numerically, these
minus signs give rise to the fermion sign problem.  With the worldline
formalism we can keep track of these permutations, and we use the
recently developed zone method~\cite{dean02} to manage the sign
problem and well as phase oscillations due to the gauge field.

\subsection{Zone method}

\begin{figure}[tbh]
\begin{center}
\includegraphics[width=6cm]{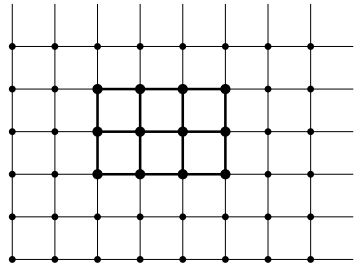}
\end{center}
\caption{Illustration of the zone method: an $6\times8$ spatial
lattice with a $3\times4$ zone.  The size of this zone is 17, as
characterized by the number of links inside the sub-lattice.
\label{fig:zone}}
\end{figure}

The zone method~\cite{dean02} consists of introducing a special
$n_{1}\times n_{2}$ spatial sub-lattice or zone.  See for example
Fig.~\ref{fig:zone}.
We allow worldline configurations which may permute fermions lying
inside this zone, but do not allow configurations that permute any
fermions lying outside of the zone.  At intermediate time slices
though the fermions are still allowed to wander through the entire
lattice.  In order to control phase oscillations associated with the
gauge field, we use a different value of the coupling $y^{\prime}$
and/or fermion mass $\mu^{\prime}$ when the fermions are outside the
zone.  We obtain physical results by extrapolation to the limit when
the zone covers the entire $L_{1}\times L_{2}$ lattice.

As demonstrated in Ref.~\cite{dean02}, observables should scale
linearly in the zone size provided that the zone size is larger than
the characteristic \textquotedblleft fermion wandering
length\textquotedblright.  For any finite values of the coupling $y$
and of the time variable $t$, this wandering length is finite, even
for massless fermions.  Thus one can extrapolate the results from
relatively small zones to the entire lattice.

\section{Numerical results}

\subsection{Zone Extrapolations}

There are different ways to define the size of a zone: by the number of
lattice points or by the number of links inside the zone.  For large
lattices it does not matter which is used.  However, for the
relatively small lattices we have used so far, it turns out that the
best way to characterize the zone size is the number of links inside
the zone.  
\begin{figure}[tbh]
{\begin{center}
\includegraphics[width=8cm]{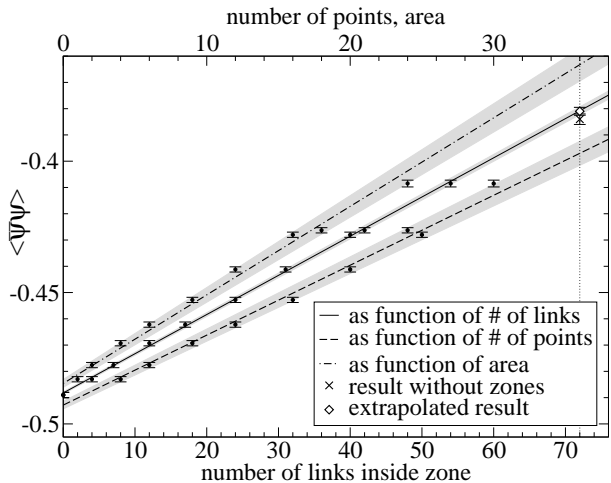}
\end{center}}
\caption{Numerical results for the condensate for different zone sizes
within a $6\times6$ lattice, with fixed values of $y=0.4$, $y'=0.1$,
$\mu=\mu'=0$, $t=1.5$, and $N=10$.  The total number of links in the
lattice is 72; the total number or points is 36, as is the total area.
The straight lines are linear fits to the data. \label{figPMlinks}}
\end{figure}
As an example, we show in Fig.~\ref{figPMlinks} the lattice condensate
for different zone sizes as function of the area of the zone, the
number of points inside a zone, and the number of links inside a zone.
A straight line fit to the condensate as function of the number of
links gives a very good fit with a $\chi^{2}$/d.o.f. of 0.6, whereas
linear fits using the number of points or the area have a
$\chi^{2}$/d.o.f. of 6.1 and 8.5 respectively.  Furthermore, the
extrapolated result, using the number of links inside a zone, does
indeed agree (within error bars) with the exact result.

To further test this method, we calculated the lattice condensate
using different parameters $y^{\prime}$ and $\mu^{\prime}$ outside the
zone, while keeping the parameters $y$ and $\mu$ inside the zone
fixed.  
\begin{figure}[tbh]
{\begin{center}
\includegraphics[width=8cm]{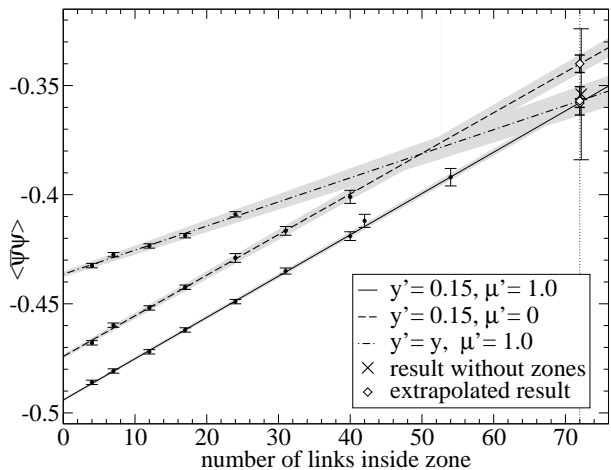}
\end{center}}
\caption{Numerical results for different zone sizes within a
$6\times6$ lattice, for fixed values of $y=0.5$, $\mu=0$, $t=1.5$, and
$N=10$.  The total number of links in the lattice is 72, where the
three linear fits meet (within enlarged error bars).
\label{figPMymuprime}}
\end{figure}
As can be seen from Fig.~\ref{figPMymuprime}, the three different sets
of data points extrapolate to results within the error bars of the
lattice condensate obtained without using the zone extrapolation
method.  The $\chi^{2}$/d.o.f. of the linear fits are 0.3, 0.7 and 1.2
respectively, indicating that the numerical data are indeed on
straight lines.

\begin{figure}[tbh]
{\begin{center}
\includegraphics[width=8cm]{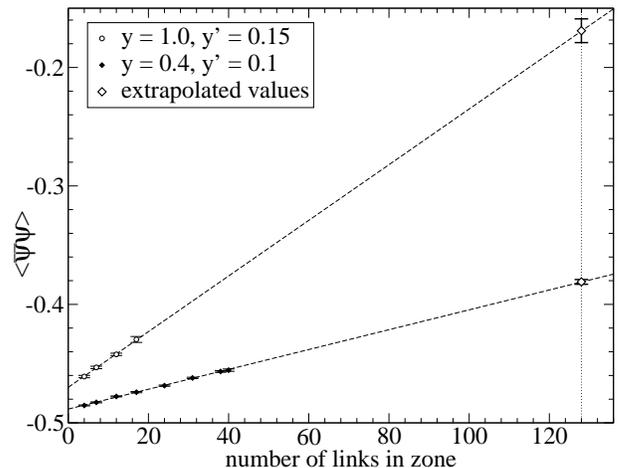}
\end{center}}
\caption{Numerical results for different zone sizes within a
$8\times8$ lattice, for values of $y=0.4$ and $y=1.0$, both with
$\mu=0$, $t=1.5$, and $N=10$.  The total number of links in the
lattice is 128. \label{figPMzone8}}
\end{figure}
Finally, in Fig.~\ref{figPMzone8} we show results on a larger lattice.
On an $8\times 8$ lattice the method seems to work quite well, although
in this case we cannot compare our result with a simulation on the
entire lattice.

\subsection{Finite size effects}

In order to avoid possible errors due to the zone extrapolation, we
checked for finite size effects without the zone method, which limits
us to rather small values of $y$ and coarse grids.  
\begin{figure}[tbh]
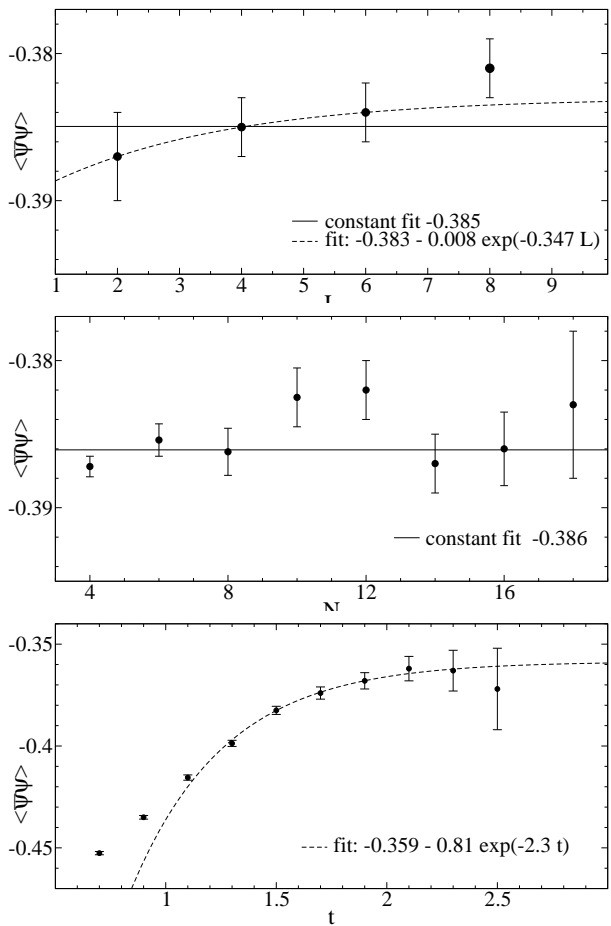

{\begin{center}
\includegraphics[width=8cm]{figPMfiniteL.eps}

\includegraphics[width=8cm]{figPMfiniten.eps}

\includegraphics[width=8cm]{figPMfinitet.eps}
\end{center}}
\caption{Finite size effects for $y=0.4$, $\mu=0$: the condensate as
function of the spatial square lattice of size $L\times L$ for $N=10$
and $t=1.5$ (top), as function of the number of time steps $N$ for
$L=4$ and $t=1.5$ (middle), and as function of $t$ for $N=10$ and
$L=4$ (bottom). \label{figPMfinite}}
\end{figure}
In Fig.~\ref{figPMfinite} we show for $y=0.4$ the dependence of the
chiral condensate on $L$, the spatial lattice size; on $N$, the number
of time steps; and on $t$, the dimensionless time variable.  We see a
slight dependence on $L$, which is actually smaller than our MC error
bars.  Note that the result for the $8\times 8$ grid was obtained
using the zone extrapolation method, where the error bar is the error
of the $\chi^{2}$ linear fit only.  Within numerical error bars, our
results are also independent of the number of time steps, $N$.

The most significant finite-size effect is the dependence on $t$, as
can be seen in the bottom panel of Fig.~\ref{figPMfinite}.  An
exponential fit of the type
\begin{eqnarray}
\langle\bar{\psi}{\psi}\rangle(t)&=&
	\langle\bar{\psi}{\psi}\rangle(t=\infty)+a_{0}\exp(-a_{1}t)
\end{eqnarray}
fits these data quite well for $t>1$.  However, for simulations at
larger lattices and larger values of $y$ the numerical errors are too
large to do a proper finite-$t$ extrapolation using such an
exponential fit.

\subsection{Summary of main results}

\begin{figure}[thb]
{\begin{center}
\includegraphics[width=8.3cm]{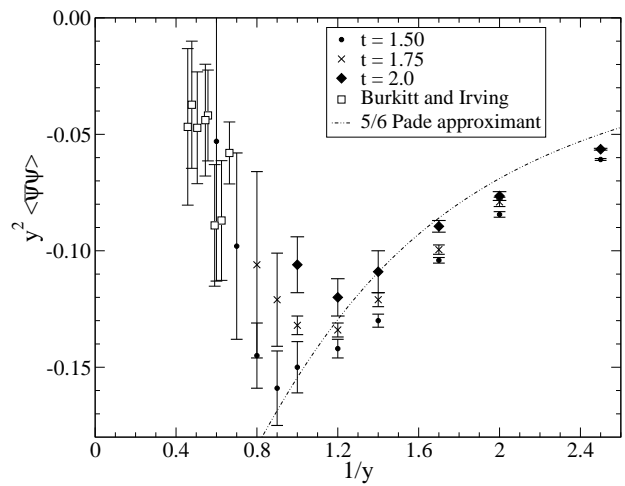}
\end{center}}
\caption{Our results for the lattice condensate as function of $y$ for
three different values of $t$ on a $6\times 6$ spatial grid, compared
to Euclidean lattice simulations data from
Ref.~\protect\cite{Burkitt:1987nx}. \label{figPMresults}}
\end{figure}

\begin{figure}[thb]
{\begin{center}
\includegraphics[width=8.3cm]{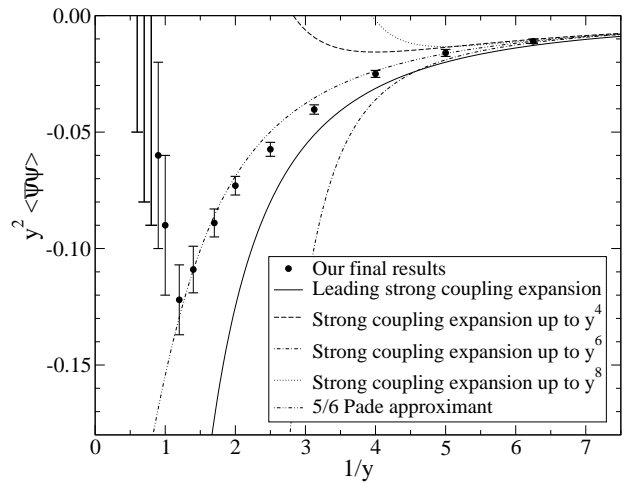}
\end{center}}
\caption{Our results for the lattice condensate as function of $y$,
obtained by extrapolating several different zone sizes for a finite
value of $t$, compared to the strong coupling
expansion~\protect\cite{Hamer:1997bf}.\label{figPMfinal}}
\end{figure}

In Figs.~\ref{figPMresults} and \ref{figPMfinal} we show our results
for a range of values of the coupling $y$.  Most of the results are
obtained on $6\times6$ spatial lattices with $N=10$ using several
different zone sizes.  The error bars in Fig.~\ref{figPMresults}
represent the $\chi^{2}$ error of the linear fit from our zone
extrapolation.  The error bars in Fig.~\ref{figPMfinal} are our best
estimate of the combined errors.  They are dominated by the $t=\infty$
extrapolation, which is based on three (or more) different values of
$t$ where possible.  For the largest values of $y$, $y\geq1.25$, we
could only establish upper limits for the condensate, due to the
uncertainties in the $t=\infty$ extrapolation.

Our results for the condensate indicate a dramatic change in behavior
around $y \sim 1$: for $y < 1$ we agree within error bars with the
$5/6$ Pad\'e approximant to the strong coupling
expansion~\cite{Hamer:1997bf}.  For $ y \to 0$ our results approach
the leading-order behavior in the strong coupling expansion
\begin{eqnarray}
 y^{2}\langle\bar{\psi}\psi\rangle^{\text{lattice}} &=& 0.5 \, y^2 \;,
\end{eqnarray}
as expected.  However, for $y>1$ we see a dramatic change in the
behavior of the condensate, and a deviation from the strong coupling
predictions: the value of the condensate decreases rapidly.  This
strong decrease of the condensate with increasing $y$ for $1<y<2$ is
in good agreement with the dynamical Euclidean Monte Carlo simulations
by Burkitt and Irving~\cite{Burkitt:1987nx}.

It is not possible to determine at this time whether or not the
condensate is small or exactly zero in the continuum limit,
$y\rightarrow \infty$.  However, it is clear from our simulations that
the continuum condensate is significantly smaller than the
prediction~\cite{Hamer:1997bf}
$y^{2}\langle\bar{\psi}\psi\rangle^{\text{lattice}} \approx 0.284$
based on the strong coupling expansion.

\section{Concluding remarks}

In this work we studied chiral symmetry breaking in one-flavor
massless QED$_{3}$, and our analysis represents the first
non-perturbative simulation of lattice gauge theory in more than one
spatial dimension with explicit fermions.  While this approach is
likely not practical for QCD in the near future, we were able to use
the zone method to control sign and phase problems to study chiral
symmetry breaking in massless QED$_{3}$.

We were able to resolve one puzzling issue regarding the size of the
chiral condensate.  Hamiltonian lattice studies had suggested a rather
large value for the chiral condensate, whereas lattice simulations and
Dyson--Schwinger studies indicated a value for the condensate about
two orders of magnitude smaller.  In our results we found that for
$y>1$ our results agree very well with the strong coupling expansion
and the condensate appears to increase as $y$ decreases.  However for
$1<y<2$ we see a rather dramatic decrease in the condensate as $y$
increases.  These results are in agreement with Euclidean lattice
simulations using staggered fermions~\cite{Burkitt:1987nx}.

In future studies we would like to study the fermion structure of the
ground state wavefunction and to compare and contrast what we see in
the simulations with the coupled cluster variational state used in
Ref.~\cite{Luo:kk}.  We also plan to study the behavior of the chiral
condensate as a function of fermion density.  We note that studies at
finite density in this Hamiltonian formalism are no more difficult
computationally than the simulations presented here.  Since Euclidean
lattice simulations of one-flavor QED$_{3}$ at finite density are also
afflicted by phase/sign oscillations (which make the computational
effort scale exponentially with volume), comparison with explicit
fermion simulations in the Hamiltonian framework could provide a
valuable numerical check.

\begin{acknowledgments}
This work was funded by the Department of Energy under Grant No.
DE-FG02-97ER41048; it benefitted from the resources of the North
Carolina Supercomputer Center.
\end{acknowledgments}


\begin{thebibliography}{99}

\bibitem {pisa}
R.~D.~Pisarski,
%``Chiral Symmetry Breaking In Three-Dimensional Electrodynamics,''
Phys.\ Rev.\ D \textbf{29}, 2423 (1984).
%%CITATION = PHRVA,D29,2423;%%

\bibitem {Burkitt:1987nx}
A.~N.~Burkitt and A.~C.~Irving,
%``Glueballs In 3-D QED With And Without Dynamical Fermions,''
Nucl.\ Phys.\ B \textbf{295}, 525 (1988).
%%CITATION = NUPHA,B295,525;%%

\bibitem{Hands:1989mv}
S.~Hands and J.~B.~Kogut,
%``Finite Size Effects And Chiral Symmetry Breaking In Quenched Three-Dimensional QED,''
Nucl.\ Phys.\ B {\bf 335}, 455 (1990).
%%CITATION = NUPHA,B335,455;%%

\bibitem{Dagotto}
E.~Dagotto, J.~B.~Kogut and A.~Kocic,
%``A Computer Simulation Of Chiral Symmetry Breaking In (2+1)-Dimensional QED With N Flavors,''
Phys.\ Rev.\ Lett.\  {\bf 62}, 1083 (1989);
%%CITATION = PRLTA,62,1083;%%
E.~Dagotto, A.~Kocic and J.~B.~Kogut,
%``Chiral Symmetry Breaking In Three-Dimensional QED With N(F) Flavors,''
Nucl.\ Phys.\ B {\bf 334}, 279 (1990).
%%CITATION = NUPHA,B334,279;%%

\bibitem {dcsb}
T.~W.~Appelquist, M.~J.~Bowick, D.~Karabali and L.~C.~Wijewardhana,
%``Spontaneous Chiral Symmetry Breaking In Three-Dimensional QED,''
Phys.\ Rev.\ D \textbf{33}, 3704 (1986);
%%CITATION = PHRVA,D33,3704;%%
T.~Appelquist, D.~Nash and L.~C.~Wijewardhana,
%``Critical Behavior In (2+1)-Dimensional QED,''
Phys.\ Rev.\ Lett.\  {\bf 60}, 2575 (1988);
%%CITATION = PRLTA,60,2575;%%
D.~Nash,
%``Higher Order Corrections In (2+1)-Dimensional QED,''
Phys.\ Rev.\ Lett.\  {\bf 62}, 3024 (1989);
%%CITATION = PRLTA,62,3024;%%
P.~Maris,
%``The influence of the full vertex and vacuum polarization on the fermion  propagator in QED3,''
Phys.\ Rev.\ D \textbf{54}, 4049 (1996).
%%CITATION = HEP-PH 9606214;%%

\bibitem{pennington}
M.~R.~Pennington and S.~P.~Webb,
%``Hierarchy Of Scales In Three-Dimensional QED,''
Brookhaven preprint BNL-40886;
%\href{http://www.slac.stanford.edu/spires/find/hep/www?r=bnl-40886}{SPIRES entry}
M.~R.~Pennington and D.~Walsh,
%``Masses From Nothing: A Nonperturbative Study Of QED In Three-Dimensions,''
Phys.\ Lett.\ B {\bf 253}, 246 (1991).
%%CITATION = PHLTA,B253,246;%%

\bibitem{Luo:kk}
X.~Q.~Luo and Q.~Z.~Chen,
%``Vacuum Structure And Chiral Symmetry Breaking In (2+1)-Dimensional Lattice Gauge Theories With Fermions,''
Phys.\ Rev.\ D {\bf 46}, 814 (1992).
%%CITATION = PHRVA,D46,814;%%

\bibitem{Azcoiti}
V.~Azcoiti and X.~Q.~Luo,
%``Phase Structure Of Compact Lattice QED In Three-Dimensions With Massless Fermions,''
Mod.\ Phys.\ Lett.\ A {\bf 8}, 3635 (1993)
[arXiv:hep-lat/9212011];
%%CITATION = HEP-LAT 9212011;%%
V.~Azcoiti, X.~Q.~Luo, C.~E.~Piedrafita, G.~Di Carlo, A.~F.~Grillo and A.~Galante,
%``Fermionic effective action and the phase structure of noncompact quantum electrodynamics in (2+1)-dimensions,''
Phys.\ Lett.\ B {\bf 313}, 180 (1993)
[arXiv:hep-lat/9305017].
%%CITATION = HEP-LAT 9305017;%%

\bibitem {Hamer:1997bf}
C.~J.~Hamer, J.~Oitmaa and Z. Weihong,
%``Series expansions for three-dimensional QED,''
Phys.\ Rev.\ D \textbf{57}, 2523 (1998).
%%CITATION = HEP-LAT 9710008;%%

\bibitem{Hands:2002dv}
S.~J.~Hands, J.~B.~Kogut and C.~G.~Strouthos,
%``Non-compact QED(3) with N(f) >= 2,''
Nucl.\ Phys.\ B {\bf 645}, 321 (2002)
[arXiv:hep-lat/0208030].
%%CITATION = HEP-LAT 0208030;%%

\bibitem {conf}
C.~J.~Burden, J.~Praschifka and C.~D.~Roberts,
%``Photon polarization tensor in three-dimensional quantum electrodynamics,''
Phys.\ Rev.\ D \textbf{46}, 2695 (1992);
%%CITATION = PHRVA,D46,2695;%%
P.~Maris,
%``Confinement and complex singularities in QED in three-dimensions,''
Phys.\ Rev.\ D \textbf{52}, 6087 (1995);
%%CITATION = HEP-PH 9508323;%%
G.~Grignani, G.~W.~Semenoff and P.~Sodano,
%``Confinement - deconfinement transition in three-dimensional QED,''
Phys.\ Rev.\ D \textbf{53}, 7157 (1996);
%%CITATION = HEP-TH 9504105;%%
M.~N.~Chernodub, E.~M.~Ilgenfritz and A.~Schiller,
%``Confinement and the photon propagator in 3D compact QED: A lattice  study in Landau gauge at zero and finite temperature,''
arXiv:hep-lat/0208013.
%%CITATION = HEP-LAT 0208013;%%

\bibitem{Dorey:1991kp}
N.~Dorey and N.~E.~Mavromatos,
%``QED in three-dimension and two-dimensional superconductivity without parity violation,''
Nucl.\ Phys.\ B {\bf 386}, 614 (1992).
%%CITATION = NUPHA,B386,614;%%

\bibitem{Farakos:1997qi}
K.~Farakos and N.~E.~Mavromatos,
%``Dynamical gauge symmetry breaking and superconductivity in  three-dimensional systems,''
Mod.\ Phys.\ Lett.\ A {\bf 13}, 1019 (1998)
[arXiv:hep-lat/9707027].
%%CITATION = HEP-LAT 9707027;%%

\bibitem{balents}
L. Balents, M. P. A. Fisher and C. Nayak, 
Phys. Rev. B \textbf{60}, 1654 (1999).

\bibitem{franz}
M. Franz and Z. Tesanovic, Phys.\ Rev.\ Lett.\ \textbf{87},
257003 (2001).

\bibitem{superTc}
M. Franz, Z. Tesanovic, and O. Vafek, 
Phys.\ Rev.\ B \textbf{65}, 180511 (2002); 
\textit{ibid.}, Phys.\ Rev.\ B \textbf{66}, 054535 (2002); 
Igor F. Herbut, Phys.\ Rev.\ Lett.\ \textbf{88}, 047006 (2002);
\textit{ibid.}, arXiv:cond-mat/0202491.

\bibitem{Maris:2002sc}
P.~Maris and D.~Lee,
%``Chiral symmetry breaking in (2+1) dimensional QED,''
arXiv:hep-lat/0209052.
%%CITATION = HEP-LAT 0209052;%%

\bibitem{dean02}
Dean Lee,
%``Permutation zones and the fermion sign problem,''
arXiv:cond-mat/0202283;
%%CITATION = COND-MAT 0202283;%%
\textit{ibid.},
%D.~Lee,
%``Zone methods and the fermion sign problem,''
arXiv:hep-lat/0209047.
%%CITATION = HEP-LAT 0209047;%%

\bibitem{Burden:qb}
C.~J.~Burden and C.~J.~Hamer,
%``Hamiltonian Strong Coupling Expansions For (2+1)-Dimensional Quantum Electrodynamics,''
Phys.\ Rev.\ D \textbf{37}, 479 (1988).
%%CITATION = PHRVA,D37,479;%%

\bibitem{worldline}
J. E. Hirsch, D. J. Scalapino, R. L. Sugar, and R. Blankenbecler, 
Phys.\ Rev.\ Lett.\ \textbf{47}, 1628 (1981); \textit{ibid.},
Phys.\ Rev.\ B \textbf{26}, 5033 (1982).

\bibitem{loop}
H. G. Evertz, G. Lana, and M. Marcu,
Phys.\ Rev.\ Lett.\ \textbf{70}, 875 (1993); 
H. G. Evertz, arXiv:cond-mat/9707221.

\bibitem{green's}
C. J. Hamer, R. J. Bursill, M. Samaras, 
Phys.\ Rev.\ D \textbf{62} 054511 (2000).

\end{thebibliography}
\end{document}